\documentclass[twocolumn,english,aps,pra,amsfonts,amsmath,amssymb,footinbib,floatfix,longbibliography]{revtex4-1}
\usepackage[latin9]{inputenc}
\usepackage{color}
\usepackage{mathrsfs}
\usepackage{amsmath}
\usepackage{amssymb}
\usepackage{graphicx}
\usepackage{esint}

\makeatletter
\usepackage{babel}
\usepackage{color}

\usepackage{filemod}

\makeatother

\usepackage{babel}
\begin{document}
\title{Analog quantum simulation of the spinor-4 Dirac equation with an artificial
gauge field}
\author{Jean Claude Garreau}
\affiliation{Universit{\'e} de Lille, CNRS, UMR 8523 -{}- PhLAM -{}- Laboratoire
de Physique des Lasers Atomes et Mol{\'e}cules, F-59000 Lille, France}
\author{V{\'e}ronique Zehnl{\'e}}
\affiliation{Universit{\'e} de Lille, CNRS, UMR 8523 -{}- PhLAM -{}- Laboratoire
de Physique des Lasers Atomes et Mol{\'e}cules, F-59000 Lille, France}
\date{\today}
\begin{abstract}
A two-dimensional spatially and temporally modulated Wannier-Stark
system of ultracold atoms in optical lattices is shown to mimic the
behavior of a Dirac particle. Suitable additional modulations generate
an artificial gauge field which simulates a magnetic field and imposes
the use of the full spinor-4 Dirac equation.
\end{abstract}
\maketitle

\section{Introduction }

The Dirac equation, unifying quantum mechanics and special relativity,
is a major achievement in physics. It has vast implications in several
fields, e.g. particle physics where it describes spin-1/2 leptons
and their interactions with photons, condensed matter where it is
used as a model for several types of quasiparticles, and also in the
fast progressing field of topological insulators.

Although the Dirac equation represents a paradigm for modern field
theory, this physics remained relatively elusive, restricted to high-energy
situations or to exotic materials. Recent developments both in condensed
matter and ultracold-atom systems have generated a burst of interest,
in particular, the concept of ``quantum simulator'' opened a new
window in the study of Dirac physics~\citep{Park:AnisotropicMasslessDirac:NP19,Garreau:SimulatingDiracModelsUltracoldBosos:PRA17,Lopez-Gonzalez:EffectiveDiracEquationOptLatt:PRA14,Qu:ObservationZitterbewegungBEC:PRA13,Tarruell:MergingDiracPtsHoneycombLattice:N12,Mazza:OpticalLatticeBasedQuantumSimulator:NJP12,Zhang:RelativisticQuantumEffectsOfDiracUltracold:FPH12,Gerritsma:QuantumSimulationKleinParadox:PRL11,Lamata:RelativisticQuantumMechanicsTrappedIons:NJP11,Gerritsma:QuantumSimulationDirac:N10,Longhi:PhotonicAnalogZitterbewegung:OL10,Dreisow:ClassicalSimulationZitterbewegungPhotLatt:PRL10,Witthaut:EffectiveDiracDynamicsBichrOptLatt:PRA11}.
Inspired by an original insight of Feynman~\citep{Feynman:SimulatingPhysics:IJTP82},
an \emph{analog} quantum simulator (as opposed to quantum-computer
simulations) is a ``simple'' and controllable system that can mimic
(some aspects of) the behavior of a more complex or less accessible
ones. The flexibility of ultracold-atom systems also prompted for
innovative ideas like the generation of the so-called ``artificial''
gauge fields acting on (neutral) atoms that mimic the effects of an
electromagnetic field~\citep{Dalibard:ArtificialGaugePotentials:RMP11},
allowing for the study of quantum magnetism~\citep{Aidelsburger:HofstadterUtracoldBosons:PRL13},
engineered spin-orbit coupling, and topological systems~\citep{Galitski:SpinOrbitCouplingQuantumGases:N13}.
The mixing of these ideas with the physics of ultracold atoms in optical
lattices has proven extremely fruitful~\citep{Bloch:ManyBodyUltracold:RMP08,Dalibard:ArtificialGaugePotentials:RMP11,Bloch:QuantumSimulationsUltracoldGases:NP14,Georgescu:QuantumSimulation:RMP14,Goldman:TopologicalQuantumMatter:NP16,Gross-Bloch:QSimulOptLatt:S17,Garreau:QuantumSimulationOfDisordered:CRP17}.

The Dirac equation is the most complete formulation describing a relativistic
charged fermion of spin 1/2. It leads to a spinor-4 description which
automatically includes the spin and the electron antiparticle, the
positron. In the rest frame of the free particle, which always exists
for a massive particle, the spinor-4 components can be separated in
two spinor-2 obeying equivalent equations. In the presence of electromagnetic
fields, however, the two spinor-2 for a massive particle generally
obey \emph{different} equations, and a full spinor-4 description is
necessary. The aim of the present work is the elaboration of a minimal
model analog to a 2D Dirac equation in the presence of an artificial
gauge field (related to a vector potential $\boldsymbol{A}$ with
non-zero rotational). In this case the spinor-4 description is mandatory,
and we show that the main characteristics of a Dirac particle are
obtained. 

In a previous work~\citep{Garreau:SimulatingDiracModelsUltracoldBosos:PRA17}
we introduced a general model allowing to mimicking a spinor-4 Dirac
Hamiltonian in 1D without magnetic field. In the present work we combine
time- and space-tailored optical potentials acting on independent
atoms in order to obtain a 2D Dirac Hamiltonian of the form 
\begin{equation}
H_{D}=c\boldsymbol{\alpha}\cdot\left(\boldsymbol{p}-q\boldsymbol{A}\right)+\beta m_{D}c^{2}\label{eq:Dirac}
\end{equation}
where $c$ is the velocity of light, $m_{D}$ and $q$ are the Dirac
particle's mass and charge, $\boldsymbol{p}$ is the momentum and
$\boldsymbol{A}$ the vector potential. The $4\times4$ Dirac matrices
constructed from Pauli matrices $\sigma_{i}$ ($i=x,y,z$) are: 
\[
\alpha_{i}=\left(\begin{array}{cc}
0 & \sigma_{i}\\
\sigma_{i} & 0
\end{array}\right)\qquad\beta=\left(\begin{array}{cc}
\boldsymbol{1} & 0\\
0 & -\boldsymbol{1}
\end{array}\right).
\]
and $\boldsymbol{1}$ is the $2\times2$ unit matrix. With this we
i) generalize this approach to the 2D case, ii) introduce an artificial
gauge field, and iii) demonstrate its ability to simulate known properties
of the Dirac equation. To the best of our knowledge, such a complete
Dirac simulator has not been proposed in the literature, and opens
new ways for a deeper exploration of the Dirac physics.

\section{Spinor-4 Dirac quantum simulator in 2D\label{sec:Spinor4}}

We first construct a modulated 2D tilted optical lattice model that
mimics a 2D spinor-4 Dirac equation with no field. The model builds
on the general approach introduced for the 1D case in Ref.~\citep{Garreau:SimulatingDiracModelsUltracoldBosos:PRA17}.
A 2D ``tilted'' (or Wannier-Stark) lattice in the $x,z$ plane~\footnote{As it will be seen in what follows, this choice makes the writing
in terms of the conventional Pauli matrices possible. We will later
introduce an artificial magnetic field in the $y$ direction. } is described by the 2D (dimensionless) Hamiltonian 
\begin{equation}
H_{0}=\frac{p_{x}^{2}+p_{z}^{2}}{2m^{*}}+V_{L}(x,z)+F_{x}x+F_{z}z\label{eq:H0}
\end{equation}
with $\boldsymbol{p}=p_{x}\mathbf{x}+p_{z}\mathbf{z}$ the momentum
in 2D ($\mathbf{x},\mathbf{z}$ are unit vectors in the directions
$x,z$), $\boldsymbol{F}=F_{x}\mathbf{x}+F_{z}\mathbf{z}$ a constant
force and $V_{L}$ a square lattice formed by orthogonal standing
waves 
\[
V_{L}(x,z)=-V_{1}\left[\cos(2\pi x)+\cos(2\pi z)\right]
\]
where space is measured in units of the step $\mathsf{a}$~\footnote{We use sans serif characters to represent dimensionfull quantities,
except when no ambiguity is possible, e.g. $\hbar$.} of the square lattice $V_{L}$, $\mathsf{a}=\lambda_{L}/2$ if the
lattice is formed by counter-propagating beams of wavelength $\lambda_{L}=2\pi/\mathrm{k}_{L}$.
Time is measured in units of $\hbar/\mathsf{E}_{R}$ where $\mathsf{E}_{R}=\hbar^{2}\mathrm{k}_{L}^{2}/2\mathrm{M}$
is the atom's recoil energy ($\mathrm{M}$ is the atom's mass). With
these choices, $m^{*}=\pi^{2}/2$ , $\hbar=1$ and $p_{j}=-i\partial_{j}$
($j=x,z)$. Since the Hamiltonian Eq.~(\ref{eq:H0}) is separable,
its eigenstates can be factorized in terms of localized Wannier-Stark
(WS) functions $\varphi_{n}^{(x,y)}$~\citep{Wannier:WS:PR60,Zener:BlochOsc:PRSA34}
which are exact eigenstates of a tilted one-dimensional lattice~\citep{Garreau:SimulatingDiracModelsUltracoldBosos:PRA17}:
\[
\Phi_{n,m}(x,z)=\varphi_{n}^{(x)}(x)\varphi_{m}^{(z)}(z)
\]
where the integer index $n$ indicates the lattice site along the
$x$ direction where the eigenfunction $\varphi_{n}^{(x)}(x)$ is
centered (resp. $m$ in the $z$ direction). The energies of the system
are then 
\begin{equation}
E_{n,m}=E_{0}+n\omega_{B}^{(x)}+m\omega_{B}^{(z)}\label{eq:ladder}
\end{equation}
where $E_{0}$ is an energy offset with respect to the bottom of the
central well $n=m=0$. The energy spacing in directions $x$ and $z$
define the so-called Bloch frequencies, $\omega_{B}^{(x)}=F_{x}$
($=\mathsf{F}_{x}\mathsf{a}/\hbar$ in dimensionfull units) and $\omega_{B}^{(z)}=F_{z}$,
where we intentionally choose $F_{z}\neq F_{x}$. This defines the
lowest ``Wannier-Stark ladder'' of energy levels separated by integer
multiples of $\omega_{B}^{(x,z)}$, Eq.~(\ref{eq:ladder}). According
to the potential parameters, there can be ``excited'' ladders, also
obeying Eq.~(\ref{eq:ladder}), but with higher values of $E_{0}$.
As in Ref.~\citep{Garreau:SimulatingDiracModelsUltracoldBosos:PRA17},
we assume here that the excited ladders are never populated and that
the dynamics occurs only among the lowest-ladder WS states of each
site.

\begin{figure*}[t]
\begin{centering}
\includegraphics[height=6cm]{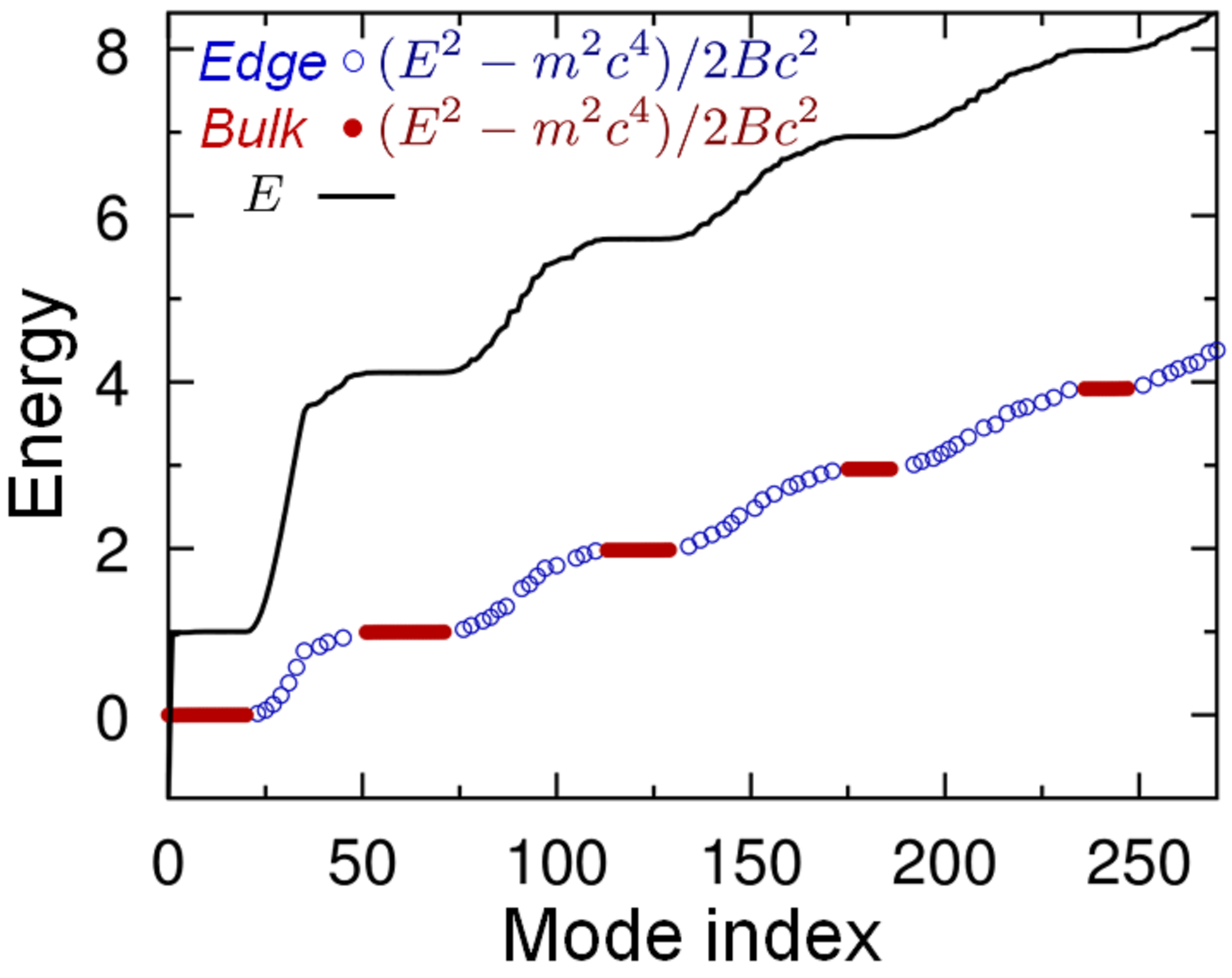}$\quad$\includegraphics[height=6cm]{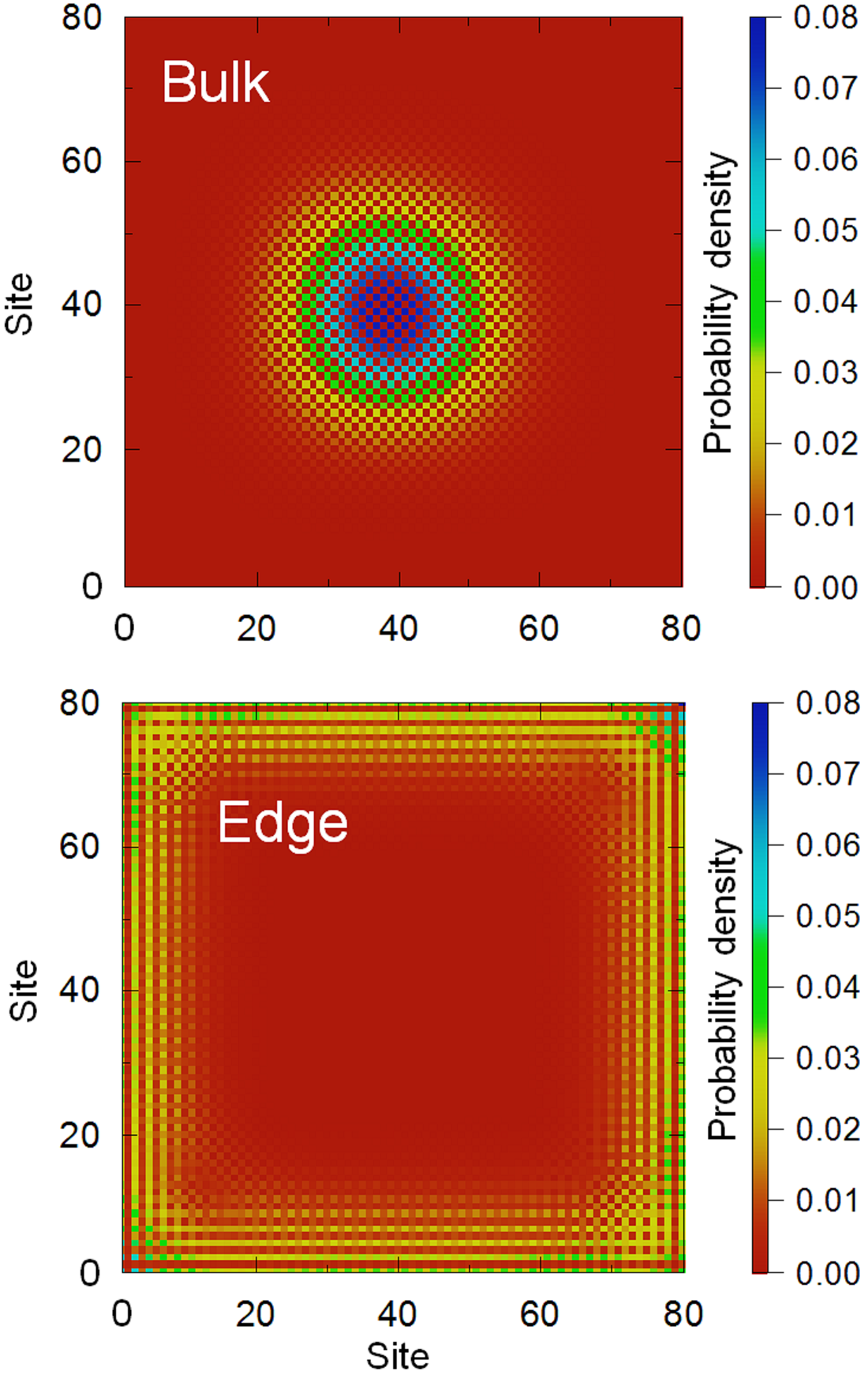} 
\par\end{centering}
\caption{\label{fig:energy}\emph{Left}: Eigenvalue spectrum of the quantum
simulator. The black solid line displays the positive raw energies
$E$ as a function of an arbitrary mode index (negative energies $E\rightarrow-E$
are not displayed for clarity). The blue circles and red disks display
the quantity $\widetilde{E}=\left(E^{2}-m^{2}c^{4}\right)/2Bc^{2}$,
Eq.~(\ref{eq:Etilde}). The red plateaus correspond to bulk states
which, according Eq.~(\ref{eq:eigen_energies}), are very close to
integers. The blue parts correspond to ``edge'' states which are
due to the finite size of the simulated lattice. \emph{Right}: Examples
of a bulk state (\emph{top}) corresponding to an integer value of
$\widetilde{E}$ and of an edge state (\emph{bottom}) corresponding
to an intermediate, non-integer value. The simulation parameters are
$mc^{2}=1$, $c_{S}=20$ and $B=1/50$, and the model parameters in
Eq.~(\ref{eq:fulldiscretemodel}) are\textcolor{blue}{{} }$T_{0}=1$,
$T_{x}=T_{y}=-10i$ , and $T_{x}^{A}=-T_{z}^{A}=1/10$.}
\end{figure*}

The Hamiltonian Eq.~(\ref{eq:H0}) is invariant under a translation
by an integer multiple $n$ of the lattice step $a=1$ (in dimensionless
units) in the $x$ direction provided that the energy is also shifted
by $nF_{x}$ (resp. $mF_{z}$ in the $z$ direction), hence the eigenstates
are such that 
\begin{equation}
\varphi_{n}^{(x)}(x)=\varphi_{0}^{(x)}(x-n),\label{eq:translation-x}
\end{equation}
(resp. $\varphi_{m}^{(z)}(z)=\varphi_{0}^{(z)}(z-m)$ in the $z$
direction).

Our aim is to obtain an effective evolution equation for the system
equivalent to a 2D Dirac equation Eq.~(\ref{eq:Dirac}) for a particle
of mass $m$ 
\begin{equation}
i\partial_{t}\left[\Psi\right]=\left\{ c\alpha_{x}\left(p_{x}-A_{x}\right)+c\alpha_{z}\left(p_{z}-A_{z}\right)+\beta mc^{2}\right\} \left[\Psi\right]\label{eq:Dirac_dimensionless}
\end{equation}
where $\partial_{t}\equiv\partial/\partial t$, $\left[\Psi\right]$
denotes a spinor-4, and $\boldsymbol{A}=A_{x}\mathbf{x}+A_{z}\mathbf{z}$.
The effective speed of light is $c=\hbar\mathrm{c}/\mathsf{a}\mathsf{E}_{R}$,
the gauge field $\boldsymbol{A}=q\mathsf{a}\mathbf{A}/\hbar$, the
mass $m=\mathsf{E}_{R}\mathsf{a}^{2}m_{D}/\hbar^{2}$, and effective
charge is set to unity. 

Consider first the case without gauge field ($\boldsymbol{A}=0$).
We can induce controlled dynamics~\citep{Garreau:SimulatingDiracModelsUltracoldBosos:PRA17}
in the system by adding to $H_{0}$, Eq.~(\ref{eq:H0}), a resonant
time-dependent perturbation of the form 
\begin{align}
\bar{V}(x,z,t)= & V_{x}\cos\left(2\pi x\right)\sin(\omega_{B}^{(x)}t)+\nonumber \\
 & V_{z}\cos\left(\pi x\right)\cos\left(2\pi z\right)\sin(\omega_{B}^{(z)}t).\label{eq:perturbation}
\end{align}
The utility of the term proportional to $\cos(\pi x)$ with spatial
periodicity $2$ will appear below. 

The general solution of the corresponding Schr\"odinger equation
can be decomposed on the eigenbasis $\Phi_{n,m}(x,z)$ :
\begin{align*}
\psi(x,z,t) & =\sum_{n,m}c_{n,m}(t)\exp\left[-i\left(n\omega_{B}^{(x)}+m\omega_{B}^{(z)}\right)t\right]\\
 & \times\varphi_{n}^{(x)}(x)\varphi_{m}^{(z)}(z)
\end{align*}
where $c_{n,m}$ are complex amplitudes. Taking into account the Hamiltonian
$H_{0}+\bar{V}$, one obtains the following set of coupled differential
equations for the amplitudes
\begin{align}
id_{t}c_{n,m}=\sum_{j,k\in\mathbb{Z}} & \left\{ \left\langle n,m\right|\bar{V}\left|n+j,m+k\right\rangle \right.\nonumber \\
 & \left.\times\exp\left[-i\left(j\omega_{B}^{(x)}+k\omega_{B}^{(z)}\right)t\right]c_{n+j,m+k}\right\} \label{eq:general coupled eq-1}
\end{align}
where $d_{t}\equiv d/dt$ and we simplified the notation $\left|\varphi_{n}^{(x)}\right\rangle \otimes\left|\varphi_{m}^{(z)}\right\rangle \equiv\left|n,m\right\rangle $.
To the dominant order, one can neglect non-resonant terms and keep
nearest neighbor couplings only, i.e take only $j=\pm1$ and $k=\pm1$
in Eq.~(\ref{eq:general coupled eq-1}) (see Appendix~\ref{app:DerivationDiracEq}).
~The time modulation at the Bloch frequencies $\omega_{B}^{(x)}$
(resp. $\omega_{B}^{(z)}$) resonantly couples first-neighbor WS states
in the $x$ direction (resp. $z$ direction). Note that as we assume
$\omega_{B}^{(x)}\neq\omega_{B}^{(z)}$ resonant couplings in both
direction $x$ and $z$ can be controlled independently.

Taking into account the choice of perturbation $\bar{V}$ of Eq.~(\ref{eq:perturbation}),
the couplings in Eq.~(\ref{eq:general coupled eq-1}) are proportional
to the spatial overlap amplitudes 
\[
\left\langle \varphi_{n}^{(x)}\right|\cos\left(2\pi x\right)\left|\varphi_{n\pm1}^{(x)}\right\rangle =\left\langle \varphi_{0}^{(x)}\right|\cos\left(2\pi x\right)\left|\varphi_{1}^{(x)}\right\rangle \equiv\Omega_{x}
\]
{[}see.~Eq.~(\ref{eq:translation-x}){]} in the $x$ direction,
and 
\begin{align*}
\left\langle \varphi_{n}^{(x)}\right|\cos\left(\pi x\right)\left|\varphi_{n}^{(x)}\right\rangle \left\langle \varphi_{m}^{(z)}\right|\cos\left(2\pi z\right)\left|\varphi_{m\pm1}^{(z)}\right\rangle \\
=\left(-1\right)^{n}\left\langle \varphi_{0}^{(x)}\right|\cos\left(\pi x\right)\left|\varphi_{0}^{(x)}\right\rangle \left\langle \varphi_{0}^{(z)}\right|\cos\left(2\pi z\right)\left|\varphi_{1}^{(z)}\right\rangle \\
=\left(-1\right)^{n}\Omega_{z}
\end{align*}
in the $z$ direction. The $(-1)^{n}$ contribution comes from the
spatial modulation in $\cos(\pi x)$ in Eq.~(\ref{eq:perturbation})
which has a period of two lattice steps {[}see Eq.~(\ref{eq:app_overlap})
in Appendix~\ref{app:DerivationDiracEq}{]} and introduces a distinction
between even and odd sites in the $x$ direction.

From Eq.~(\ref{eq:general coupled eq-1}), one obtains a set of differential
equations with nearest-neighbor couplings.
\begin{align}
id_{t}c_{n,m}= & T_{x}\left(c_{n+1,m}-c_{n-1,m}\right)\nonumber \\
 & +(-1)^{n}T_{z}\left(c_{n,m+1}-c_{n,m-1}\right)\label{eq:cnm_sans_masse}
\end{align}
where $T_{x}=-iV_{x}\Omega_{x}/2$ and $T_{z}=-iV_{z}\Omega_{z}/2$.

Assuming that the wave packet is large and smooth at the scale of
the lattice step one can take the continuous limit of Eq.~(\ref{eq:cnm_sans_masse}),
and associate the discrete amplitudes $c_{n,m}(t)$ to two continuous
functions. Since the site parity dependence in Eq.~(\ref{eq:cnm_sans_masse})
must be taken into account, we introduce the functions $s_{ee}(x,z,t)$
which is the envelope of $c_{n,m}$ for $n,m$ even, $s_{eo}(x,z,t)$
which is the envelope of $c_{n,m}$ for $n$ even and $m$ odd, and
analogously for $s_{oe}$ and $s_{oo}$. These functions can be arranged
as components of a spinor-4 $[\Psi]=\left(s_{ee},s_{oo},s_{eo},s_{oe}\right){}^{\intercal}$
describing 4 coupled sub-lattices, which, from Eq.~(\ref{eq:cnm_sans_masse}),
obey a Hamiltonian equation 
\begin{equation}
id_{t}\left[\Psi(x,z,t)\right]=H_{S}\left[\Psi(x,z,t)\right]\label{eq:Dirac-1}
\end{equation}
where $H_{S}$ can be easily shown to be analogous to a 2D Dirac Hamiltonian
of the form $H_{S}=c_{S}\mathbf{\boldsymbol{\alpha}}\cdot\boldsymbol{p}$
with an effective velocity of light $c_{S}=V_{x}\Omega_{x}=V_{z}\Omega_{z}$
(these terms can be made equal by adequately tuning the modulation
amplitudes $V_{x,z}$, but one can also create an anisotropic model
with $c_{S_{x}}\neq c_{S_{z}}$, with the possibility of an effective
violation of Lorentz invariance\textcolor{blue}{~\citep{Park:AnisotropicMasslessDirac:NP19}}).
The above system can be broken into two equivalent set of equations~\citep{Garreau:SimulatingDiracModelsUltracoldBosos:PRA17}
which correspond to the well-known massless \textit{\emph{Weyl spinor-2
fermion}}. The twofold-degenerated dispersion relation deduced from
Eq.~(\ref{eq:Dirac-1}) is $\omega(k)=\pm\left(\Omega_{x}^{2}k_{x}^{2}+\Omega_{z}^{2}k_{z}^{2}\right)^{1/2}$
and corresponds, as it could be expected, to a Dirac cone.

The Hamiltonian of a massive particle can be generated by adding a
static perturbation $V_{0}\cos(\pi x)\cos(\pi z)$ to Eq.~(\ref{eq:perturbation}).
Then, neglecting derivatives $\partial_{x,z}^{2}c_{m,n}$ or higher,
Eq.~(\ref{eq:cnm_sans_masse}) takes the form 
\begin{align}
id_{t}c_{n,m}= & (-1)^{n+m}T_{0}+(-1)^{n}T_{z}\left(c_{n,m+1}-c_{n,m-1}\right)\nonumber \\
 & +T_{x}\left(c_{n+1,m}-c_{n-1,m}\right)\label{eq:discretemass}
\end{align}
where $T_{0}=V_{0}\left\langle \varphi_{0}^{(z)}\right|\cos(\pi z)\left|\varphi_{0}^{(z)}\right\rangle \left\langle \varphi_{0}^{(x)}\right|\cos(\pi x)\left|\varphi_{0}^{(x)}\right\rangle $.
In the continuous limit one obtains a 2D Dirac Hamiltonian for a finite-mass
particle $H_{S}=c_{S}\mathbf{\boldsymbol{\alpha}\cdot p}+\beta mc^{2}$
with an effective rest energy $mc^{2}=T_{0}$. This term opens a gap
of width $2mc^{2}$ in the the dispersion relation, separating ``particle''
and ``antiparticle'' states, exactly as in the Dirac equation.

The approximations used in construction of the above model are well
understood and controlled. As the system is to a very good approximation
a closed one (even experimentally), there is only a small broadening
of the resonances, which make these resonances highly selective. Next-to-neighbor
couplings are out of resonance and are thus negligible, and the same
is true for intrawell couplings between different ladders. For particular
values of the potential parameters $(V_{x,z},F_{x,z})$ ``accidental''
resonances can occur, but these exotic situations are not considered
here. The accuracy of our approach can also be checked by comparing
the results of the model to a simulation of the full Schr\"odinger
equation, which takes into account all existing couplings, as we have
done with the simpler, but conceptually equivalent, 1D model in Ref.~\citep{Garreau:SimulatingDiracModelsUltracoldBosos:PRA17}.
For the very same reason (the sharpness of the resonances) the system
is not expected to be particularly sensitive to heating caused by
experimental noise.

\section{Dirac equation with an artificial gauge field \label{sec:Dirac-with_gauge} }

In this section, we show how the analogous of a ``non-trivial''
(i.e. of non-zero rotational) vector potential $\boldsymbol{A}=A_{x}\mathbf{x}+A_{z}\mathbf{z}$
with $A_{x}\propto z$ and $A_{z}\propto x$, can introduced in the
system. The Dirac Hamiltonian of Eq.~(\ref{eq:Dirac}) can be realized
by adding to Eq.~(\ref{eq:perturbation}) an additional perturbation
\begin{align}
\bar{V}_{A}(x,z,t)= & V_{x}^{A}z\cos\left(2\pi x\right)\cos(\omega_{B}^{(x)}t)+\nonumber \\
 & V_{z}^{A}x\cos\left(\pi x\right)\cos\left(2\pi z\right)\cos(\omega_{B}^{(z)}t).\label{eq:perturbation-1}
\end{align}
This term can be treated in the same way as in Sec.~\ref{sec:Spinor4}
and leads to 
\begin{align}
id_{t}c_{n,m} & =(-1)^{n+m}T_{0}+(-1)^{n}T_{z}\left(c_{n,m+1}-c_{n,m-1}\right)\nonumber \\
 & +T_{x}\left(c_{n+1,m}-c_{n-1,m}\right)\nonumber \\
 & +T_{x}^{A}m\left(c_{n+1,m}+c_{n-1,m}\right)\nonumber \\
 & +(-1)^{n}nT_{z}^{A}\left(c_{n,m+1}+c_{n,m-1}\right)\label{eq:fulldiscretemodel}
\end{align}
where $T_{z}^{A}=\frac{1}{2}\Omega_{z}V_{z}^{A}$, $T_{x}^{A}=\frac{1}{2}\Omega_{x}V_{x}^{A}$.
The additional terms on the last two lines in Eq.~(\ref{eq:fulldiscretemodel})
are due to the potential Eq.~(\ref{eq:perturbation-1}) which generates
the slopes proportional to $m$ (resp. $n$) in direction $z$ (resp.
$x$)~\footnote{Experimentally, such linear terms can be obtained by superposing two
laser fields of frequencies separated by $\delta\omega$, which generate
a beat note varying in space as $\sin\left(\delta\omega x/c\right)\sim\delta\omega x/c$;
if $c/\delta\omega$ is large enough compared to the lattice wavelength
the linear approximation can be valid over a large number of neighbor
sites where the wavepacket is concentrated.}.

In the continuous limit, and neglecting second order and higher derivatives
of the spinor components, we obtain the Hamiltonian Eq.~(\ref{eq:Dirac}),
with an artificial gauge field $\boldsymbol{A}=-2(zT_{x}^{A}\mathbf{x}+xT_{z}^{A}\mathbf{z})/c_{S}$
(cf.~App.~\ref{app:DerivationDiracEq}). The symmetric gauge considered
in the next section can be realized by tuning the modulation amplitudes
in $\bar{V}_{A}$ so that $T_{x}^{A}=-T_{z}^{A}$, so that $\boldsymbol{A}=2T_{z}^{A}(z\mathbf{x}-x\mathbf{z})/c_{S}$,
corresponding to a uniform magnetic field in the $y$ direction $\boldsymbol{B}=4T_{z}^{A}\mathbf{y}/c_{S}$~\footnote{The Landau gauge can be obtained by setting $T_{z}^{A}=0$ (or $T_{x}^{A}=0$).}.

\begin{figure*}
\begin{centering}
\includegraphics[height=4cm]{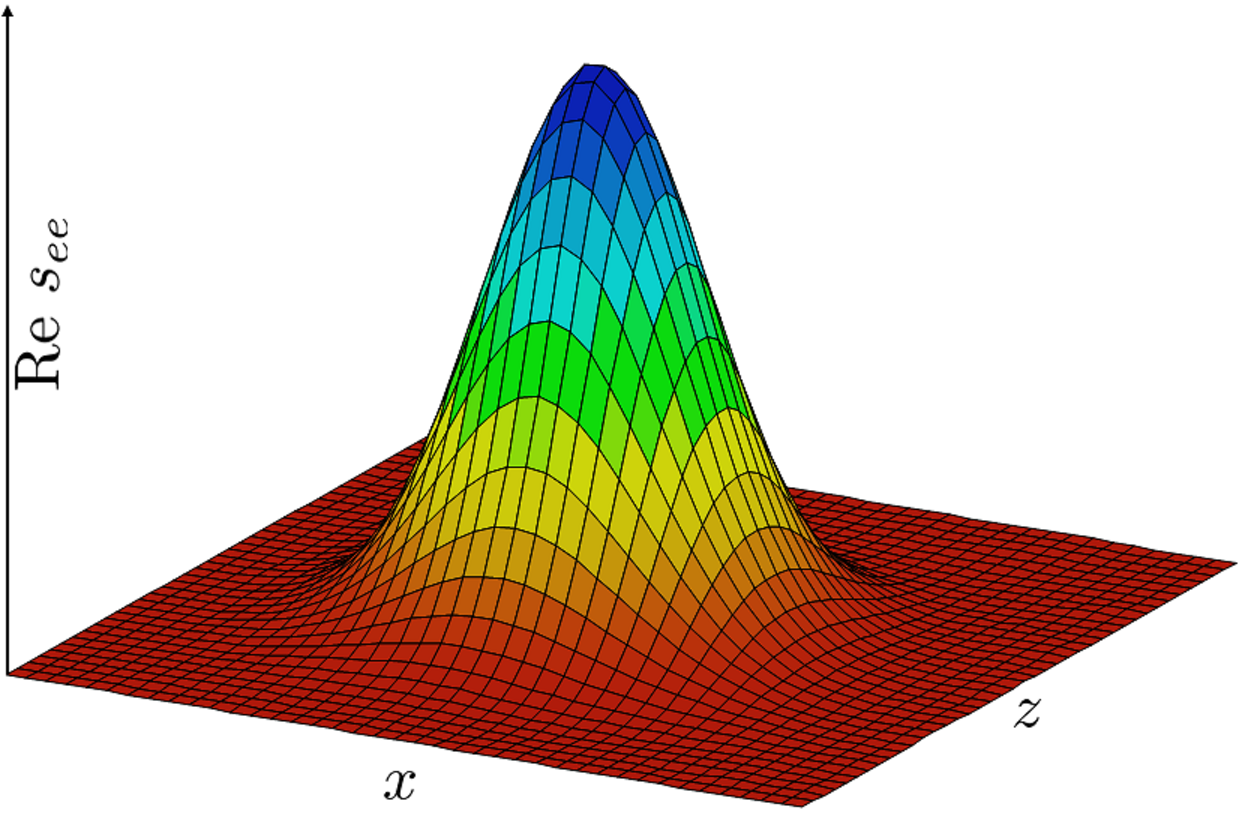}$\;$\includegraphics[height=4.2cm]{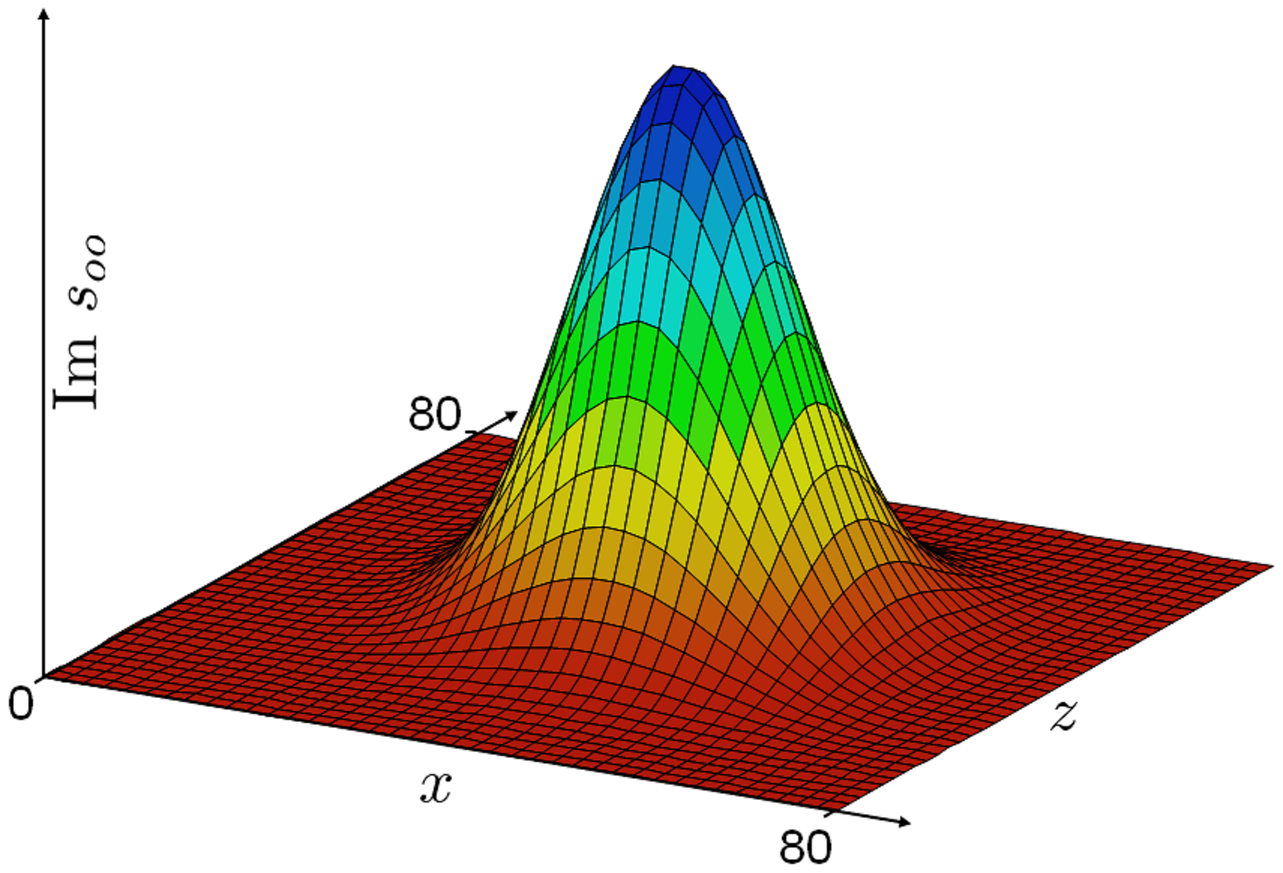} 
\par\end{centering}
\centering{}\includegraphics[height=4cm]{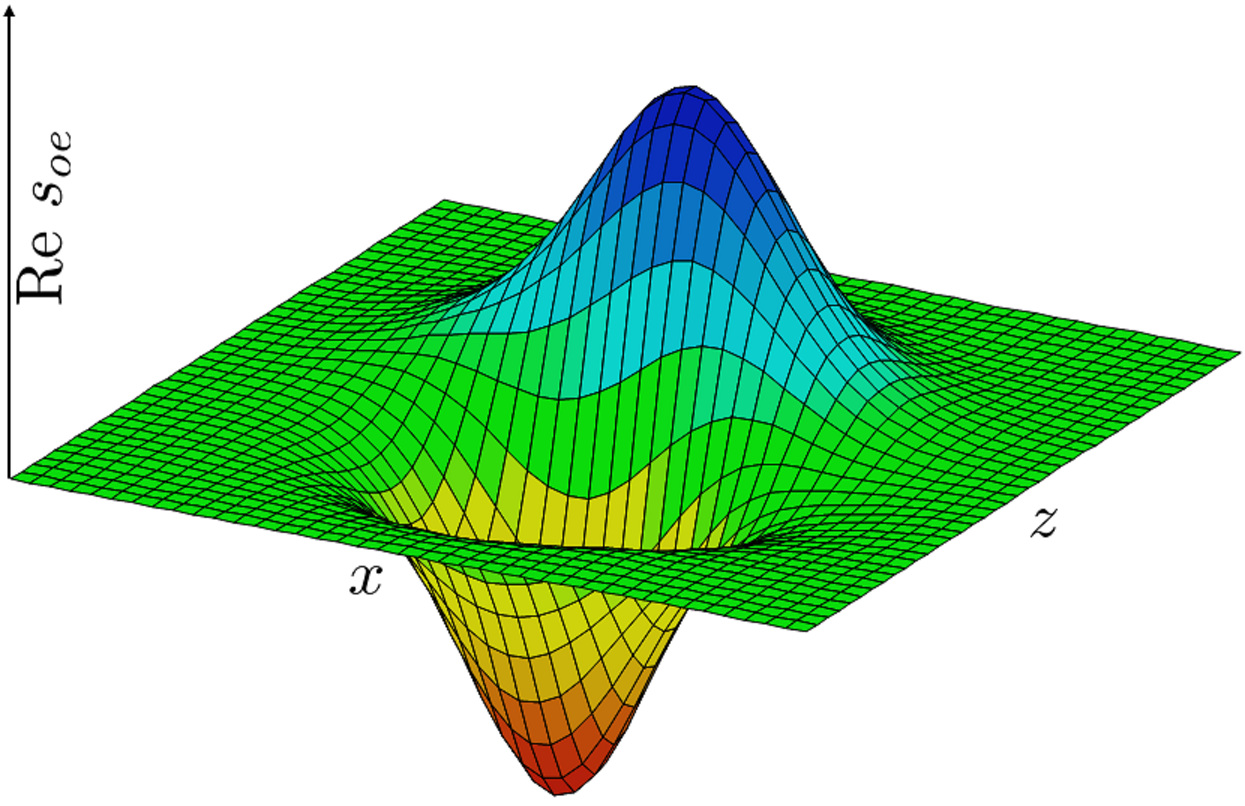}$\;$\includegraphics[height=4cm]{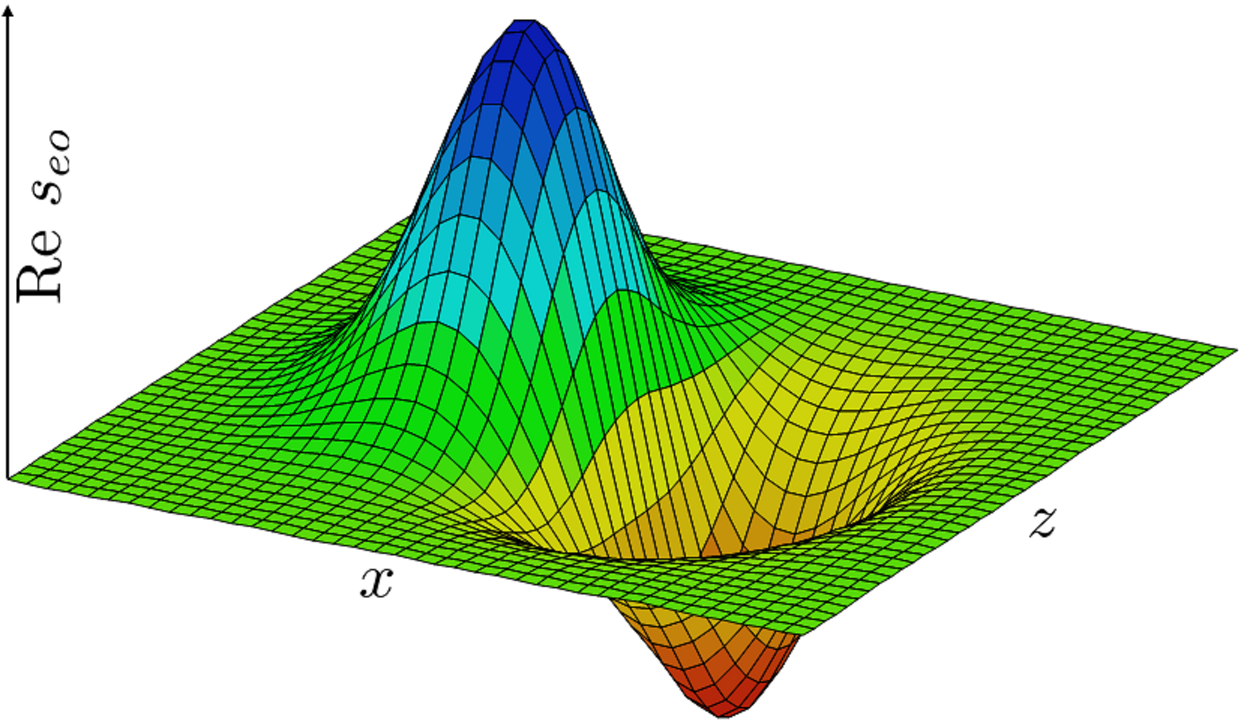}\caption{\label{fig:excit}First ``bulk'' excited eigenstate of Eq.~(\ref{eq:mode_excit}).
Four components of the spinor are shown: Re($s_{ee}$), Im($s_{oo}$),
Re($s_{oe}$), Re($s_{eo}$). The first excited energy is $E=4.11$
(same parameters as in Fig.~ \ref{fig:energy}). The WS numerical
simulation was performed in a square of $80\times80$ lattice sites
(shown in only one plot, for clarity). }
\end{figure*}

\section{Quantum simulation of Dirac physics\label{sec:SimulationDiracPhysics}}

We briefly recall well-known theoretical results of Eq.~(\ref{eq:Dirac_dimensionless})
for a spinor-4 Dirac particle in the presence of a symmetric vector
potential $\boldsymbol{A}=2T_{z}^{A}(z\mathbf{x}-x\mathbf{z})/c_{S}$.
We will compare them with the numerical results of the discrete model
Eq.~(\ref{eq:fulldiscretemodel}) and show that they are in very
good agreement, proving that the discrete model reproduces to a good
level of accuracy the Dirac physics.

In the following, we write the spinor-4 as $\left[\Psi\right]=\left(\begin{array}{c}
\phi\\
\chi
\end{array}\right)$ where $\phi,\chi$ are spinor-2s. With this convention, the Dirac
equation can be decomposed in 2 coupled equations 
\begin{align}
(E-mc^{2})\phi & =c\left[\boldsymbol{\sigma}\cdot\left(\boldsymbol{p}-\boldsymbol{A}\right)\right]\chi\label{eq:spinor_phi_chi}\\
(E+mc^{2})\chi & =c\left[\boldsymbol{\sigma}\cdot\left(\boldsymbol{p}-\boldsymbol{A}\right)\right]\phi.\label{eq:spinor_chi_phi}
\end{align}
These equations are symmetric under the transposition $E,$$\phi$,
$\chi$ $\longleftrightarrow$ $-E,$$-\chi$, $\phi$ so that the
negative energy states are easily deduced from their positive energy
counterparts.

Eliminating $\chi$ from Eqs.~(\ref{eq:spinor_chi_phi}) gives $(E^{2}-m^{2}c^{4})\phi=c\left[\boldsymbol{\sigma}\cdot\left(\boldsymbol{p}-\boldsymbol{A}\right)\right]^{2}\phi$,
and, after some straightforward algebra, one finds

\begin{widetext}
\begin{align}
(E^{2}-m^{2}c^{4})\phi= & c^{2}\left[\left(p^{2}+\frac{B^{2}}{4}(x^{2}+z^{2})+L_{y}B\right)\mathbf{1}+\sigma_{y}B\right]\phi\label{eq:spinor-2}
\end{align}
\end{widetext}where $L_{y}=(zp_{x}-xp_{z})$ is the angular momentum
component in the $y$ direction, with a ``diamagnetic'' term proportional
to $x^{2}+z^{2}$ and ``paramagnetic'' terms of the type $\boldsymbol{L}\cdot\boldsymbol{B}$
and $\boldsymbol{\sigma}\cdot\boldsymbol{B}$. This equation has ``spin
up'' $\phi_{+}$ and ``spin down'' $\phi_{-}$ spinor-2~\footnote{We label these solutions ``spin up'' and ``spin down'' (with quotation
marks) because they are proportional to the eigenstates $(1,\pm i)^{\intercal}/\sqrt{2}$
of $\sigma_{y}$ (remember that $\boldsymbol{B}\parallel\mathbf{y}$).
This is not to be confused with spinor-4-particle's spin components.} solutions 
\begin{equation}
\phi_{\pm}=(1,\pm i)^{\intercal}\psi_{\pm}(x,z)\label{eq:sinor_phi}
\end{equation}
where the functions $\psi_{\pm}(x,z)$ obey the following equation

\begin{equation}
\left[\frac{p^{2}}{2\mu}+\frac{\mu}{2}(x^{2}+z^{2})+L_{y}\right]\psi_{\pm}=(\widetilde{E}\mp1)\psi_{\pm}\label{eq:HarmonicOsc}
\end{equation}
with 
\begin{equation}
\widetilde{E}=(E^{2}-m^{2}c^{4})/(Bc^{2})\label{eq:Etilde}
\end{equation}
and $\mu=B/2$, which strongly evokes a 2D harmonic oscillator of
mass $\mu$ and natural frequency $\omega=1$ in a magnetic field.

\begin{figure*}
\centering{}\includegraphics[height=4cm]{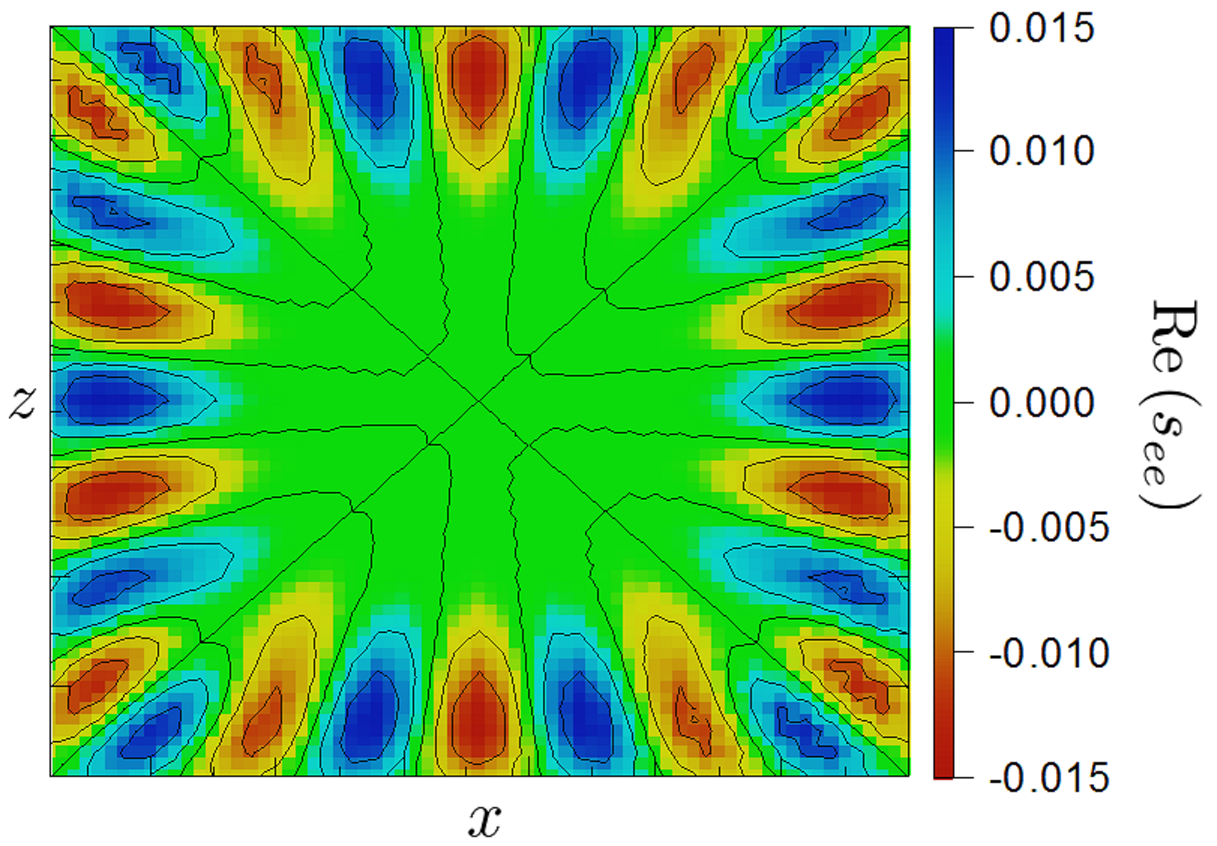}$\;$\includegraphics[height=4cm]{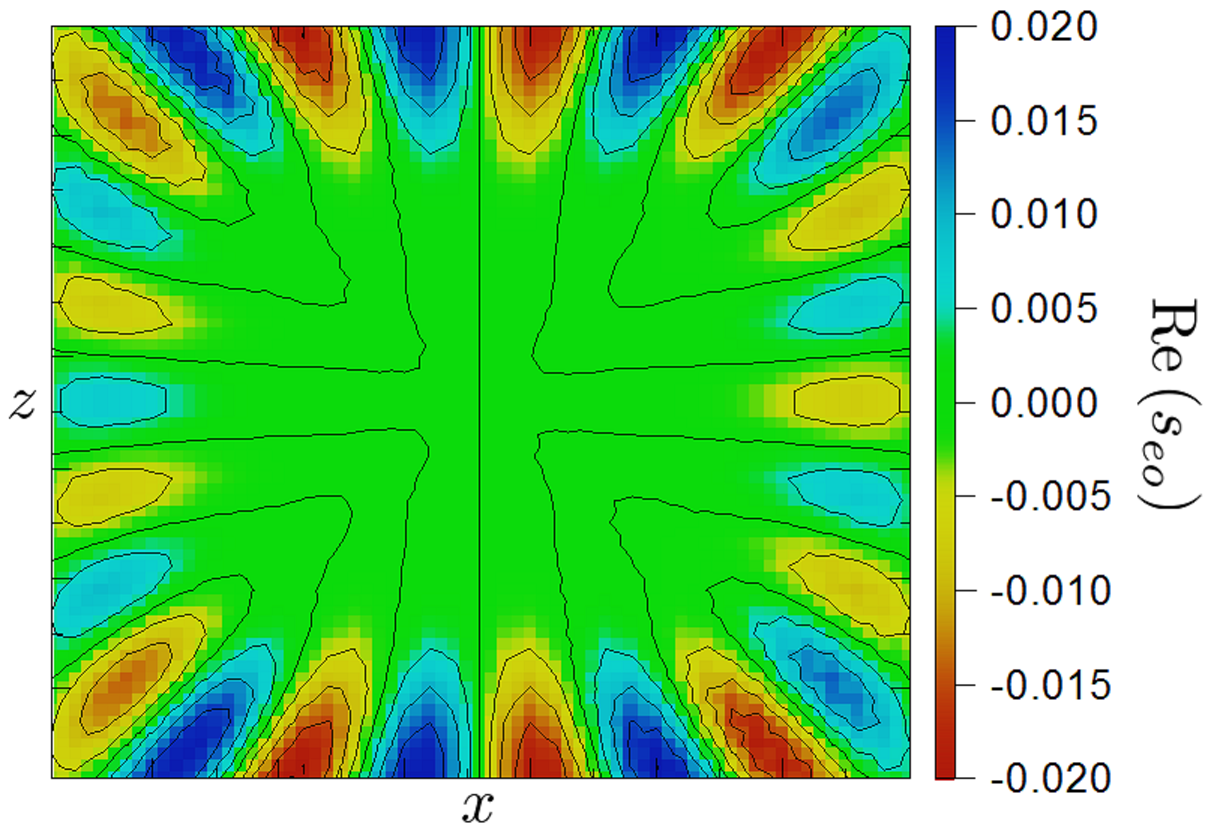}\\
\includegraphics[height=4cm]{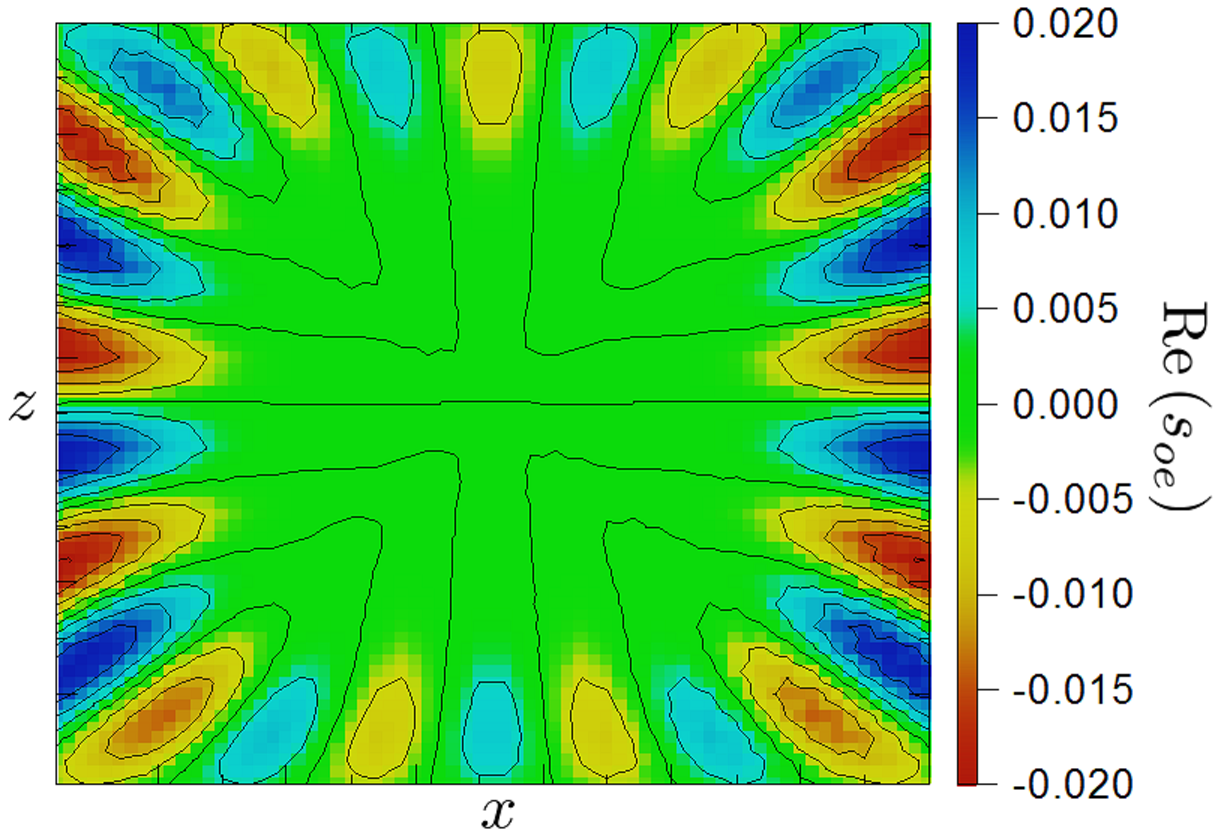}$\;$\includegraphics[height=4cm]{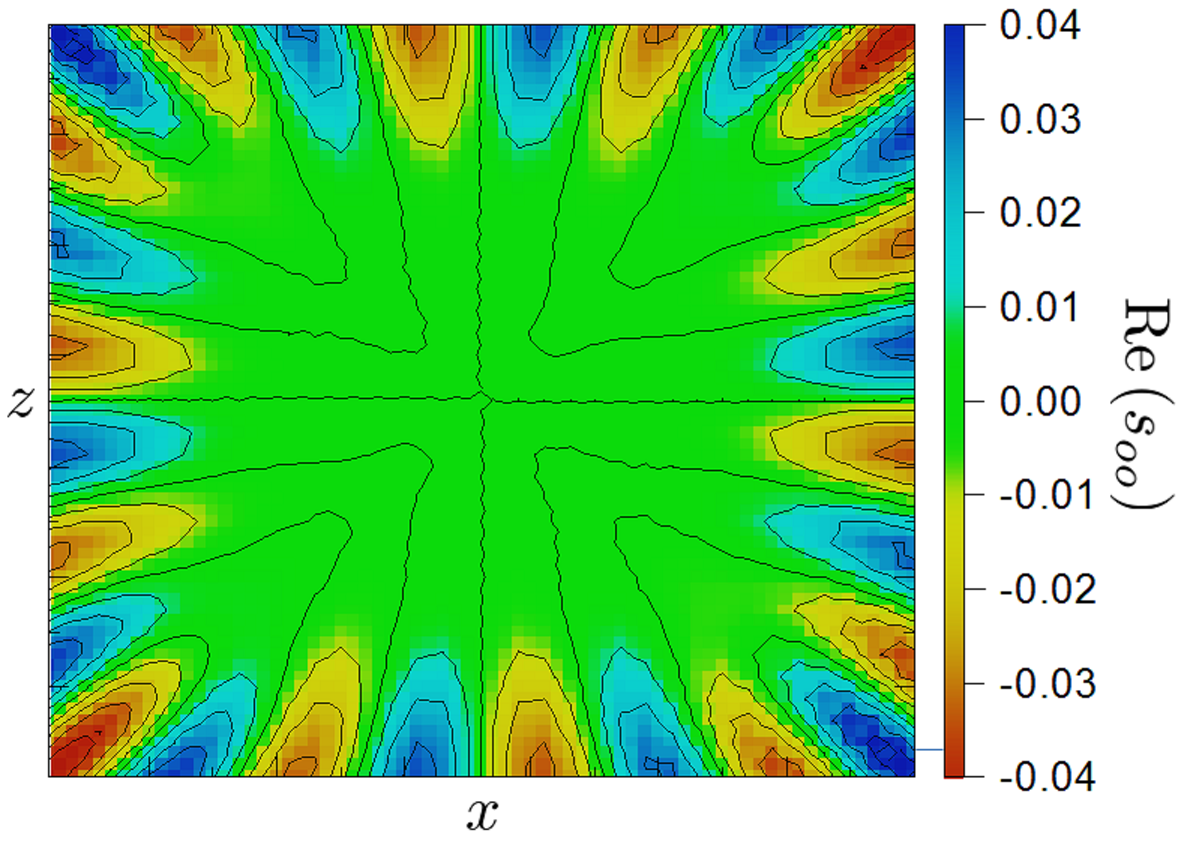}\caption{\label{fig:Borderstate}Chiral border state. The real part of the
spinor-4 components $s_{ee},s_{eo},s_{oe},s_{oo}$ clearly display
a wave propagating along the borders. The imaginary parts are analogous,
but dephased of $\pi/2$.}
\end{figure*}

Solutions of Eq.~(\ref{eq:HarmonicOsc}) are the well-known ``Landau
levels'', with spectrum 
\begin{equation}
E=\sqrt{m^{2}c^{4}+2Bc^{2}n}\label{eq:eigen_energies}
\end{equation}
where $n$ (not to be confused with the site index $n$, which does
not appear in the present section) is a strictly positive natural
number for ``spin up'' solution $\phi_{+}$ and a natural number
for ``spin down'' $\phi_{-}$. For each $n$ the corresponding energy
is infinitely degenerated.

The solutions of Eq.~(\ref{eq:HarmonicOsc}) are also well known,
and together with Eq.~(\ref{eq:sinor_phi}) and Eq.~(\ref{eq:spinor_chi_phi})
lead to solutions for the full eigenspinors $\left[\Psi\right]$\textcolor{blue}{.}
In polar coordinates $x=r\cos\varphi$, $z=r\sin\varphi$ and $\rho=r/\ell$,
where $\ell=\sqrt{2/B}$ is an effective dimensionless ``magnetic
length'', an example of solution for the fundamental level $E=mc^{2}$
is 
\[
\left(\begin{array}{c}
\phi\\
\chi
\end{array}\right)=\mathscr{N}\left(\begin{array}{c}
(1,-i)^{\intercal}\exp(-\rho^{2}/2)\\
(0,0)^{\intercal}
\end{array}\right)
\]
where $\mathscr{N}$ is a normalization factor. There is a corresponding
negative energy $E=-mc^{2}$ counterpart with $\phi\rightarrow\phi$
and $\chi\rightarrow-\chi$. Examples of eigenstates belonging to
the $n=1$ excited state family $E=\left(m^{2}c^{4}+2Bc^{2}\right)^{1/2}$
\begin{equation}
\left(\begin{array}{c}
\phi\\
\chi
\end{array}\right)=\mathscr{N}\left(\begin{array}{c}
(1,i)^{\intercal}\\
(1,-i)^{\intercal}\beta\rho\exp(-i\varphi)
\end{array}\right)\exp(-\rho^{2}/2)\label{eq:mode_excit}
\end{equation}
or
\[
\left(\begin{array}{c}
\phi\\
\chi
\end{array}\right)=\mathscr{N}\left(\begin{array}{c}
(1,-i)^{\intercal}\rho\exp(-i\varphi)\\
(1,i)^{\intercal}\beta
\end{array}\right)\exp(-\rho^{2}/2)
\]
where $\beta=-2c/\left[\ell(mc^{2}+E)\right]$~\footnote{Note that these states are also eigenstates of the angular momentum
operator $L_{y}+\varSigma_{y}$ , where $\varSigma_{y}=\frac{1}{2}\left(\begin{array}{cc}
\sigma_{y} & 0\\
0 & \sigma_{y}
\end{array}\right)$ is the spin operator. }.

Numerical simulations of the discrete model reproduce the characteristics
of the above solutions of the Dirac equation. According to Eq.~(\ref{eq:eigen_energies}),
plateaus are obtained at integer values $n=\left(E^{2}-m^{2}c^{4}\right)/2Bc^{2}$.
As shown in Fig.~\ref{fig:energy}, the numerical solution of Eq.~(\ref{eq:fulldiscretemodel})
displays such a behavior, with plateaus (red disks) appearing as expected
at integer values. The simulation also shows eigenenergies which fall
at non-integer values (blue circles); we checked that these are edge
states due to finite size effects, as the numerical simulation is
performed in a square box containing finite number of lattice sites
(80$\times$80)~\footnote{We numerically asserted that a $150\times150$ lattice gives essentially
the same results.}.

\begin{figure}
\begin{centering}
\includegraphics[height=4cm]{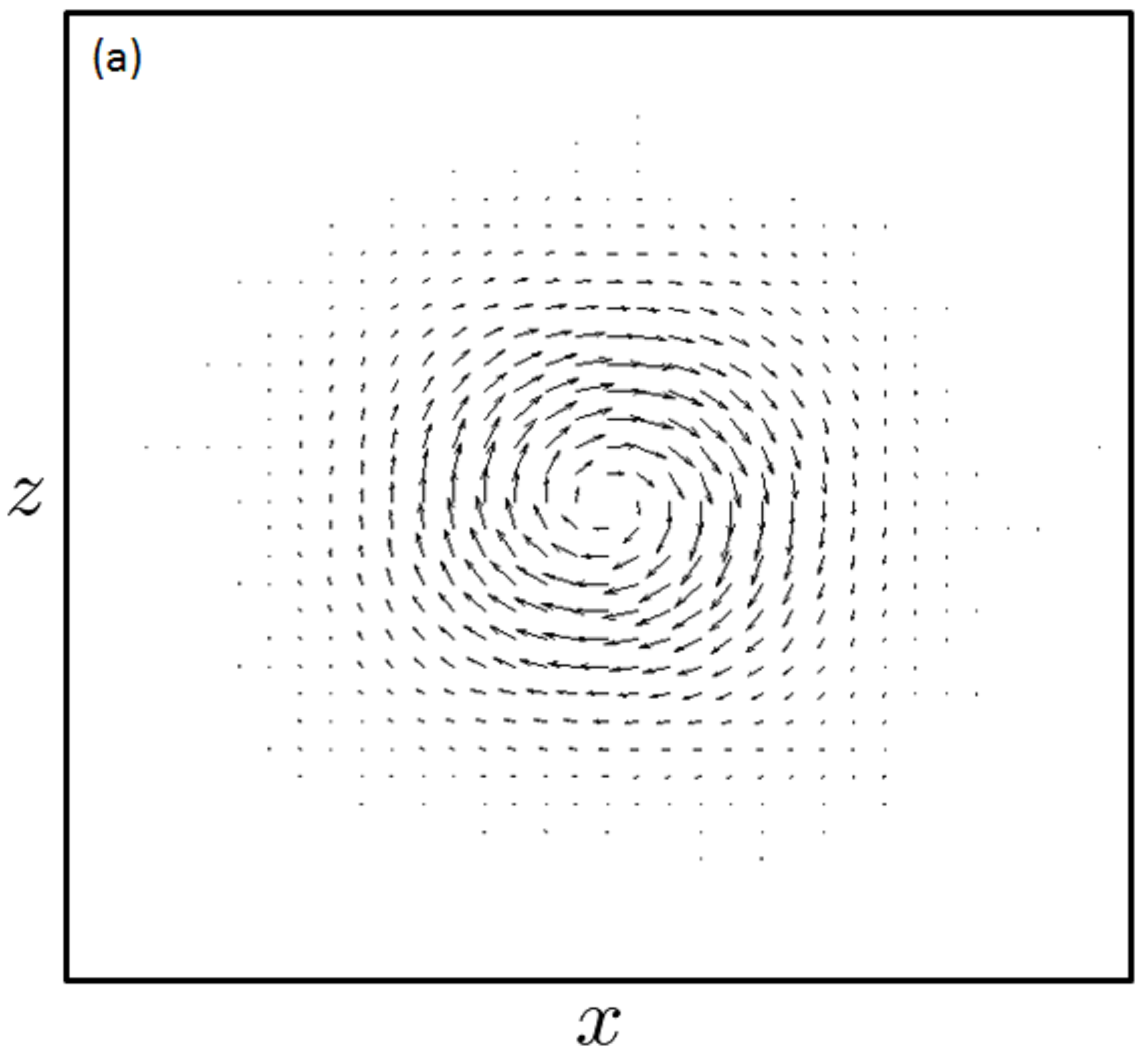}$\;$\includegraphics[height=4cm]{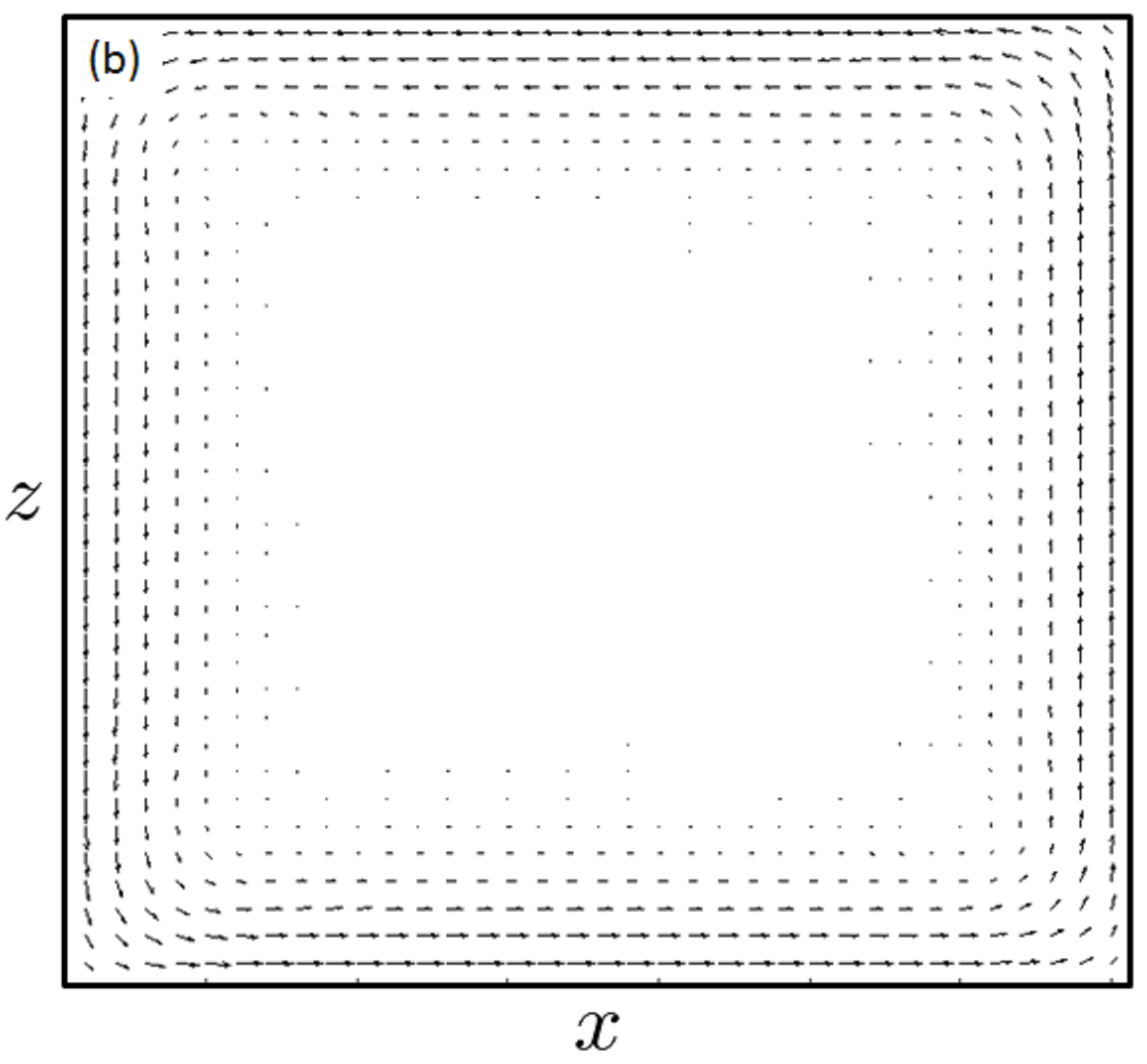}
\par\end{centering}
\caption{\label{fig:current}Chiral current vector field for (a) the bulk state
in Fig.~\ref{fig:excit} Eq.~(\ref{eq:mode_excit}) and (b) the
border state in Fig.~\ref{fig:Borderstate}. }
\end{figure}

Figure~\ref{fig:excit} shows a numerical example of the the four
components of the eigenspinor $\left[\Psi\right]=\left[s_{ee},s_{oo},s_{eo},s_{oe}\right]^{\intercal}$
for the bulk state, corresponding to the analytical solution of the
Dirac excited mode of Eq.~(\ref{eq:mode_excit}); injecting the magnetic
length $\ell=10$ in Eq.~(\ref{eq:mode_excit}), we get a quantitative
agreement with the simulation result. This confirms the ability of
our model to mimic the behavior of a spinor-4 Dirac particle. Figure~\ref{fig:Borderstate}
is the false-colors representation of the border state shown in Fig.~\ref{fig:energy},
clearly displaying its chiral edge-state nature which evokes Quantum
Hall Effect states. Recently, the existence of chiral Quantum-Hall-Effect-like
states in acoustic systems received great attention~\citep{Peri:AxialFieldInducedChiralChannels:NP19,Wen:AcousticLandauQuantizationQHE:NP19},
opening prospects for the simulation of Dirac physics in such systems.
However, in contrast with our spinor-4 simulator, these acoustic simulators
are described by a Weyl equation for a spinor-2. Finally, Fig.~\ref{fig:current}
display the chiral vector fields of the true \emph{Dirac spinor-4
current of probability} for the bulk and for the edge states (see
App.~\ref{sec:DiracCurrent}), further evidencing the ability of
our system to reproduce spinor-4 Dirac physics.

\section{Conclusion}

The present work demonstrates the quantum simulation a Dirac particle
in the presence of a magnetic field, putting together quantum simulations
of ``exotic'' dynamics (from the point of view of low-energy physics)
and artificial gauge fields. This is intrinsically expected to lead
to topological systems, as the Dirac cone is one of the main (and
simplest) examples of topology in physics, and opens new ways to explore
exciting new possibilities in this very active field. Other interesting
prospects would be the inclusion of particle-particle interactions
and/or disorder in this analog quantum simulations. These are intriguing
and highly challenging tasks, beyond the scope of the present work.\medskip{}

\begin{acknowledgments}
This work was supported by Agence Nationale de la Recherche through
Research Grants K-BEC No. ANR-13-BS04-0001-01 and MANYLOK No. ANR-18-CE30-0017,
the Labex CEMPI (Grant No. ANR-11-LABX-0007-01), the Ministry of Higher
Education and Research, Hauts-de-France Council and European Regional
Development Fund (ERDF) through the Contrat de Projets Etat-Region
(CPER Photonics for Society, P4S). 
\end{acknowledgments}

\appendix

\section{Derivation of the Dirac equation}

\label{app:DerivationDiracEq}

This appendix discusses in more detail the developments leading to
Eqs.~(\ref{eq:cnm_sans_masse}) and (\ref{eq:discretemass}) for
the site amplitudes $c_{n,m}$ and explains how the equations for
their continuous envelopes can be put in an form analogous to the
Dirac equation. The derivation of the Dirac equation in presence of
a gauge field of Sec.~\ref{sec:Dirac-with_gauge} follows the same
steps and is also sketched below.

Injecting the general wave function $\psi(x,z,t)=\sum_{n,m}c_{n,m}(t)\exp\left(-i(n\omega_{B}^{(x)}+m\omega_{B}^{(z)})t\right)\varphi_{n}^{(x)}\varphi_{m}^{(z)}$
in the Schr\"odinger equation for the Hamiltonian $H=H_{0}+\widetilde{V}$
Eq.~(\ref{eq:perturbation}), one obtains the following set of coupled
differential equations
\begin{align}
id_{t}c_{n,m}=\sum_{j,k\in\mathbb{Z}} & \left\{ \left\langle n,m\right|\widetilde{V}\left|n+j,m+k\right\rangle \right.\nonumber \\
 & \left.\times\exp\left[-i\left(j\omega_{B}^{(x)}+k\omega_{B}^{(z)}\right)t\right]c_{n+j,m+k}\right\} .\label{eq:general coupled eq}
\end{align}
To the dominant order, neglecting non-resonant terms, and keeping
nearest neighbor couplings~\footnote{This is justified by the rapid vanishing of overlapping integrals
between WS states localized in different sites.} only $j,k=\pm1$ in Eq.~(\ref{eq:general coupled eq}) one gets

\begin{widetext}

\begin{eqnarray}
id_{t}c_{n,m} & = & -i\frac{V_{x}}{2}\left(\left\langle n,m\right|\cos(2\pi x)\left|n+1,m\right\rangle c_{n+1,m}-\left\langle n,m\right|\cos(2\pi x)\left|n-1,m\right\rangle c_{n-1,m}\right)\nonumber \\
 &  & -i\frac{V_{z}}{2}\left(\left\langle n,m\right|\cos(\pi x)\cos(2\pi z)\left|n,m+1\right\rangle c_{n,m+1}-\left\langle n,m\right|\cos(\pi x)\cos(2\pi z)\left|n,m-1\right\rangle c_{n,m-1}\right).\label{eq:Evolutioncnm}
\end{eqnarray}
The overlap integrals can be simplified using the Wannier-Stark state
translation properties as in Sec.~\ref{sec:Spinor4}. For instance:

\begin{align}
\left\langle n,m\right|\cos(\pi x)\cos(2\pi z)\left|n,m\pm1\right\rangle  & =\intop dx\varphi_{n}^{(x)}(x)\varphi_{n}^{(x)}(x)\cos(\pi x)\intop dz\varphi_{m}^{(z)}(x)\varphi_{m\pm1}^{(z)}(x)\cos(2\pi z)\nonumber \\
 & =\intop dx\varphi_{0}^{(x)}(x-n)\varphi_{0}^{(x)}(x-n)\cos(\pi x)\intop dz\varphi_{0}^{(z)}(x-m)\varphi_{\pm1}^{(z)}(x-m)\cos(2\pi z)\nonumber \\
 & =(-1)^{n}\intop dx\varphi_{0}^{(x)}(x)\varphi_{0}^{(x)}(x)\cos(\pi x)\intop dz\varphi_{0}^{(z)}(x)\varphi_{1}^{(z)}(x)\cos(2\pi z)\nonumber \\
 & =(-1)^{n}\Omega_{z}\label{eq:app_overlap}
\end{align}
defining the coupling parameter $\Omega_{z}$ {[}the $(-1)^{n}$ factor
come from the perturbation in $\cos(\pi x)${]}. The same calculation
can be done for overlaps $\left\langle n,m\right|\cos(2\pi x)\left|n\pm1,m\right\rangle $
and gives

\[
\left\langle n,m\right|\cos(2\pi x)\left|n\pm1,m\right\rangle =\intop dx\varphi_{0}^{(x)}(x)\varphi_{1}^{(x)}(x)\cos(2\pi x)=\Omega_{x}.
\]
\end{widetext}These results lead to Eq.~(\ref{eq:cnm_sans_masse}). 

When a static perturbation $V_{0}\cos(\pi x)\cos(\pi z)$ is added
to Eq.~(\ref{eq:perturbation}), we have, in the resonant approximation,
an additional term corresponding to a ``self-coupling'' contribution
\[
V_{0}\left\langle n,m\right|\cos(\pi x)\cos(\pi z)\left|n,m\right\rangle c_{n,m}
\]
which is to be added to the R.H.S. of Eq.~(\ref{eq:cnm_sans_masse}).
The overlap integral is then

\begin{widetext}
\begin{eqnarray*}
\left\langle n,m\right|\cos(\pi x)\cos(\pi z)\left|n,m\right\rangle  & = & \intop dx\varphi_{n}^{(x)}(x)\varphi_{n}^{(x)}(x)\cos(\pi x)\intop dz\varphi_{n}^{(z)}(x)\varphi_{n}^{(z)}(x)\cos(\pi x)\\
 & = & (-1)^{n+m}\intop dx\varphi_{0}^{(x)}(x)\varphi_{0}^{(x)}(x)\cos(\pi x)\intop dz\varphi_{0}^{(z)}(x)\varphi_{0}^{(z)}(x)\cos(\pi x)
\end{eqnarray*}
\end{widetext}which in turn gives Eq.~(\ref{eq:discretemass}).

In taking the continuous limit, we have to account for the factors
$(-1)^{n}$ and $(-1)^{n+m}$ in Eq.~(\ref{eq:cnm_sans_masse}) or
Eq.~(\ref{eq:discretemass}) which lead to parity-dependent smooth
envelopes. Taking as an example $n,m$ even, Eq.~(\ref{eq:Evolutioncnm})
shows that $c_{n,m}$ is coupled to the next-neighbors amplitudes
with $n$ and $m$ of opposite parities, i.e. $(n\pm1,m)$ or $(n,m\pm1)$
\begin{align*}
id_{t}c_{n,m}= & T_{0}c_{n,m}+T_{x}\left(c_{n+1,m}-c_{n-1,m}\right)\\
 & +T_{z}\left(c_{n,m+1}-c_{n,m-1}\right),\qquad n,m\;\:\mathrm{even}
\end{align*}
with the coupling constants $T_{j}=-iV_{j}\Omega_{j}/2$ ($j=x,z$)
defined as in Sec.~\ref{sec:Spinor4}. Taking the continuous limit
$c_{n+1,m}-c_{n-1,m}\simeq2\partial_{x}c_{oe}$ and $c_{n,m+1}-c_{n,m-1}\simeq2\partial_{z}c_{eo}$,
one has
\begin{eqnarray*}
id_{t}s_{ee}(x,z,t) & = & T_{0}s_{ee}(x,z,t)+2T_{x}\partial_{x}s_{oe}(x,z,t)\\
 &  & +2T_{z}\partial_{x}s_{eo}(x,z,t)\\
 & = & T_{0}s_{ee}(x,z,t)+\Omega_{x}p_{x}s_{oe}(x,z,t)\\
 &  & +\Omega_{z}p_{z}s_{eo}(x,z,t)
\end{eqnarray*}
where $p_{x}=-i\partial_{x}$, $p_{z}=-i\partial_{z}$. The calculation
is analogous for the remaining smooth components $s_{oo},$ $s_{oe}$
and $s_{eo}$.

The introduction of the artificial gauge field in Sec.~\ref{sec:Dirac-with_gauge}
follows the same lines as above. The additional modulation terms {[}see
Eq.~(\ref{eq:perturbation-1}){]} generate extra terms in the evolution
equation for the amplitudes $c_{n,m}.$ For instance, the term proportional
to $V_{x}^{A}z\cos\left(2\pi x\right)\cos(\omega_{B}^{(x)}t)$ gives
in the R.H.S. of the evolution equation for amplitudes $c_{n,m}$
the following contribution:\begin{widetext}
\begin{align*}
\frac{V_{x}^{A}}{2}\left(\left\langle n,m\right|z\cos(2\pi x)\left|n+1,m\right\rangle c_{n+1,m}+\left\langle n,m\right|z\cos(2\pi x)\left|n-1,m\right\rangle c_{n-1,m}\right)\\
=\frac{V_{x}^{A}}{2}\left\langle n,m\right|z\cos(2\pi x)\left|n+1,m\right\rangle \left(c_{n+1,m}+c_{n-1,m}\right)
\end{align*}
\end{widetext}with overlap integrals that can then be written as
\begin{align*}
\left\langle n,m\right|z\cos(2\pi x)\left|n\pm1,m\right\rangle  & =\Omega_{x}\left\langle m\right|z\left|m\right\rangle \\
 & =\Omega_{x}\intop dzz\left|\varphi_{m}(z)\right|^{2}\\
 & =\Omega_{x}\intop dz(z+m)\left|\varphi_{0}(z)\right|^{2}\\
 & =\Omega_{x}\left\langle \bar{z}+m\right\rangle 
\end{align*}
where the constant $\bar{z}$ is a small offset which can be canceled
by a $z$ translation. Equation~(\ref{eq:fulldiscretemodel}) is
then easily obtained, as well as its continuous limit for the envelopes
of $c_{ab}$ ($a,b=e,o$), assuming they are smooth enough to allow
neglecting second order derivatives (for instance $c_{n+1,m}+c_{n-1,m}\simeq2s_{ee}(x,z,t)$
for $n$ odd and $m$ even, and so on) .

\section{Dirac current }

\label{sec:DiracCurrent}

The Dirac equation leads to a standard probability current continuity
equation $\partial_{t}\left(\Psi^{\dagger}\Psi\right)+\mathbf{\boldsymbol{\nabla}}\cdot\boldsymbol{j}=0$
with the following current components
\[
j_{i}=\left[\Psi\right]^{\dagger}\alpha_{i}\left[\Psi\right]=\phi^{\dagger}\sigma_{i}\chi+\chi^{\dagger}\sigma_{i}\phi
\]
where $\left[\Psi\right]=\left[\phi;\chi\right]$ as in Sec.~\ref{sec:SimulationDiracPhysics}.
For instance, taking the of Eq.~(\ref{eq:mode_excit}) the current
is
\[
j_{x}\propto-\rho e^{-\rho^{2}}\sin\varphi,\;j_{z}=-\rho e^{-\rho^{2}}\cos\varphi,
\]
which is in excellent agreement with Fig.~\ref{fig:current}(a).


%

\end{document}